\documentclass[aps,pre,twocolumn]{revtex4-1}
\usepackage{amsmath,bm,epsfig,,subfigure}
\usepackage{color}
\usepackage[normalem]{ulem}
\def\Fbox#1{\vskip1ex\hbox to 8.5cm{\hfil\fboxsep0.3cm\fbox{%
  \parbox{8.0cm}{#1}}\hfil}\vskip1ex\noindent}  

\newcommand{\B}[1]{{\bm{#1}}}
\let \= \equiv \let\*\cdot \let\~\widetilde \let\-\overline

\def\d{{\rm d}}

\begin{document}
\title{Plasticity-Induced Magnetization in Amorphous Magnetic Solids}
\author{H. George E. Hentschel, Itamar Procaccia and Bhaskar Sen Gupta}
\affiliation{Dept of Chemical Physics, The Weizmann Institute of Science, Rehovot 76100, Israel}
\begin{abstract}
Amorphous magnetic solids, like metallic glasses, exhibit a novel effect: the growth of magnetic order as a function of mechanical strain under athermal conditions in the presence of a magnetic field. The magnetic moment increases in steps whenever there is a plastic event. Thus plasticity induces the magnetic ordering, acting as the effective noise driving the system towards equilibrium. We present results of atomistic simulations of this effect in a model of a magnetic amorphous solid subjected to pure shear and a magnetic field. To elucidate the dependence on external strain and magnetic field we offer a mean-field theory that provides an adequate qualitative understanding of the observed phenomenon.
\end{abstract}
\maketitle

\section{Introduction}

The study of the influence of random impurities, disorder and random anisotropy on the magnetic properties of crystalline
materials is an often studied and therefore an extremely well established subject of condensed matter physics \cite{73HPZ,75IM,93Seth}. In contrast, the theoretical study of the consequences of glassy randomness on the interaction between mechanical and magnetic properties in amorphous solids like metallic glasses is still in its infancy. In this paper we study an interesting effect that has been discovered in our group via numerical simulations; an amorphous
magnetic solid with random local anisotropy quenched from the liquid in the presence of a magnetic field has an initial  magnetic moment (which is zero in the absence of magnetic field), cf. Fig. \ref{numeffect}. But when subjected to an external mechanical strain the magnetic moment can increase with the accumulation of plastic events. The magnetization increases to a steady state level that depends on the relative
magnitude of the magnetic field compared to the other parameters in the problem. Needless to say, this effect is particular to glassy systems in which the position of particles is free to adjust under mechanical strains and plastic irreversibility; its study requires models that combine glassiness, magnetism and randomness.

In Sect. \ref{model} we present the model that we use to study the interaction between mechanical and magnetic responses
in amorphous solids \cite{12HIP,13DHPS}. In Sect. \ref{effect} we present the numerical results of the effect under discussion. We show that when the magnetic field is non-zero, the magnetization increases to a steady-state value that
depends on the magnetic field. In Sect. \ref{ss} we offer a theory for the steady-state value of the magnetization. In
 Sect. \ref{mvsg} we present a mean field theory for the actual trajectory of the magnetization as a function of
 external strain, and present some non-trivial predictions of the theory. Sect. \ref{conclude} contains
a summary of the paper and some concluding remarks.

\section{The Model}
\label{model}

The model we employ is in the spirit of the Harris, Plischke and Zuckerman (HPZ) Hamiltonian \cite{73HPZ}
but with a number of important modifications to conform with the physics of amorphous magnetic solids \cite{12HIP}. First, our particles are not pinned to a lattice.  We write the Hamiltonian as
\begin{equation}
\label{umech}
U(\{\B r_i\},\{\B S_i\}) = U_{\rm mech}(\{\B r_i\}) + U_{\rm mag}(\{\B r_i\},\{\B S_i\})\ ,
\end{equation}
where $\{\B r_i\}_{i=1}^N$ are the 2-D positions of $N$ particles in an area $L^2$ and $\B S_i$ are spin variables. The mechanical part $U_{\rm mech}$ is chosen to represent a glassy material with a binary mixture of 65\% particles A and 35\% particles B,
with Lennard-Jones potentials having a minimum at positions $\sigma_{AA}=1.17557$, $\sigma_{AB}=1.0$ and $\sigma_{BB}=0.618034$ for
the corresponding interacting particles \cite{09BSPK}. These values are chosen to guarantee good glass formation and avoidance of crystallization. The energy parameters chosen are $\epsilon_{AA}=\epsilon_{BB}=0.5$
$\epsilon_{AB}=1.0$, in units for which the Boltzmann constant equals unity. All the potentials are truncated at distance 2.5$\sigma$ with two continuous derivatives. Particles A carry spins $\B S_i$; the B particles are not magnetic.
We choose the spins $\B S_i$ to be
classical $xy$ spins; the orientation of each spin is then given by an angle $\phi_i$. We also denote by $\theta_i({\bf r}_i)$ the local preferred easy axis of anisotropy, and end up with the magnetic contribution to the potential energy in the form \cite{12HIP}:
\begin{eqnarray}
&&U_{\rm mag}(\{\B r_i\}, \{\B S_i\}) = - \sum_{<ij>}J(r_{ij}) \cos{(\phi_i-\phi_j)}\nonumber\\&&-  \sum_i K_i\cos^2{(\phi_i-\theta_i(\{\B r_i\}))}-  \mu_A B \sum_i \cos{(\phi_i)} \ .
\label{magU}
\end{eqnarray}
Here $r_{ij}\equiv |\B r_i-\B r_j|$ and the sums are only over the A particles that carry spins. For a discussion of the physical significance of each term the reader is referred to Ref.~\cite{12HIP}. It is important however to stress that in our model (in contradistinction with the HPZ Hamiltonian \cite{73HPZ} and also with the Random Field Ising Model \cite{93Seth}), the exchange parameter $J(\B r_{ij})$ is a function of a changing inter-particle position (either due
to external strain or due to non-affine particle displacements, and see below). Thus randomness in the exchange interaction is coming from the random positions $\{\B r_i\}$, whereas the function $J(\B r_{ij})$ is not random. We choose for concreteness the monotonically decreasing form $J(x) =J_0 f(x)$ where $f(x) \equiv \exp(-x^2/0.28)+H_0+H_2 x^2+H_4 x^4 $ with
$H_0=-5.51\times 10^{-8}\ ,H_2=1.68 \times 10^{-8}\ , H_4=-1.29 \times 10^{-9}$.
This choice cuts off $J(x)$ at $x=2.5$ with two smooth derivatives.  In our case $J_0=1$.

Another important difference is that in our case
the local axis of anisotropy $\theta_i$ is {\em not} selected from a pre-determined distribution, but is determined by the local structure: define  the matrix $\B T_i$:
\begin{equation}
T_i^{\alpha\beta} \equiv \sum_j J( r_{ij})  r_{ij}^\alpha r_{ij}^\beta/\sum_j J( r_{ij}) \ .
\end{equation}
The matrix $\B T_i$ has two eigenvalues in 2-dimensions that we denote as $\kappa_{i,1}$ and $\kappa_{i,2}$, $\kappa_{i,1}\ge \kappa_{i,2}$. The eigenvector that belongs to the larger eigenvalue $\kappa_{i,1}$ is denoted by $\hat {\B n}$. The easy axis of anisotropy is given by by $\theta_i\equiv \sin^{-1} (|\hat n_y|)$. Finally the coefficient $K_i$ which now changes from particle to particle is defined as
\begin{equation}
\label{KK}
K_i \equiv \tilde C[\sum_j J( r_{ij})]^2 (\kappa_{i,1}-\kappa_{i,2})^2\ ,~~ \tilde C= K/J_0\sigma^4_{AB} \ .
\end{equation}
The parameter $K$ determines the strength of this random local anisotropy term compared
 to other terms in the Hamiltonian. The form given by Eq.~(\ref{KK}) ensures that for an isotropic distribution of particles $K_i=0$. Due to the glassy random nature of our material the direction $\theta_i$ is random. In fact we will assume below (as can be easily tested in the numerical simulations) that the angles $\theta_i$ are distributed randomly in the interval $[-\pi,\pi]$. It is important to note that external straining does NOT change this flat distribution and we will assert that the probability distribution $P(\theta_i)$ can be simply taken as
\begin{equation}
P(\theta_i)d\theta_i = \frac{d\theta_i}{2\pi} \ .
\label{ptheta}
\end{equation}
The last term in Eq. (\ref{magU}) is
the interaction with the external field $B$. We have chosen $\mu_A B$ in the range [-0.08,0.08]. At the two extreme values all the spins are aligned along the direction of $\B B$.

\section{Plasticity Induced Magnetization}
\label{effect}

The novel effect that is the subject of this paper is described as follows: we prepare the system described
by the Hamiltonian (\ref{umech}) in a fluid state at a high temperature ($T=1.2$ in units of
$\epsilon_{AB}$ where the Boltzmann constant is fixed at unity). In this paper the system is 2-dimensional, containing $N=2000$ particles.  The system is then quenched to $T=0$ with molecular dynamics, and finally brought to an inherent state using gradient energy minimization \cite{99ML,04ML,09LP} {\em in the presence of a magnetic field $B$ directed in the $x$ direction }. All the subsequent simulation are performed at $T=0$. We strain the system
using simple shear, such that at each step of strain $\delta \gamma$ the particles are first subjected to an
affine transformation
\begin{equation}
x_i\to x_i+\delta \gamma y_i\ , \quad y_i\to y_i \ ,
\end{equation}
where $\delta\gamma=0.001$. Subsequent to the affine step the system is again relaxed to equilibrium using energy
gradient minimization \cite{99ML,04ML,09LP}. This procedure of athermal, quasi-static strain (AQS) is continued until we reach the desired values
of the strain $\gamma$. The procedure is applied either in the absence of magnetic field ($B=0$) or at any chosen
value of $\B B$ which is always directed in the $x$ direction.

\begin{figure}
\includegraphics[scale = 0.35]{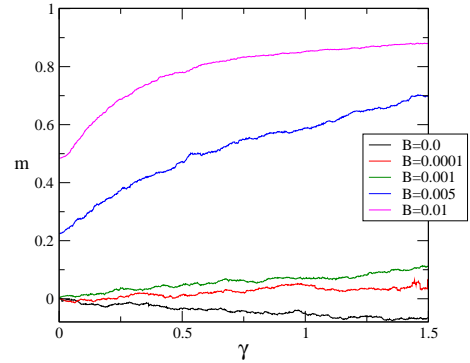}
\caption{The magnetization $m$ as a function of the strain $\gamma$ for different values of the magnetic field
$B$ as shown in the inset. Note that each value of $B$ is a associated with an initial value $m(B,\gamma=0)$ and a steady state value $m_{ss}(B)$. We
will develop a theory below for both the steady state value and the trajectory to get there.}
\label{numeffect}
\end{figure}

Immediately after the quench from the liquid at $\gamma=0$ the magnetic moment $m$ has a value that depends on the
magnetic field $B$, $m(B,\gamma=0)$. Denoting the number
of spin carrying particles as $N_A$ we write quite generally
\begin{equation}
m\equiv \langle \cos \phi\rangle =\frac{1}{N_A}\sum_i \cos \phi_i \ .
\label{defm}
\end{equation}
The effect of interest here is what happens to $m$ as we begin to strain the system in the AQS procedure. The
answer is shown in Fig. \ref{numeffect} which exhibits $m$ as a function of $\gamma$ for $\gamma\le 1.5$ for various values of the magnetic field $\B B$. For $\gamma\to \infty$ (not shown in the figure) the magnetization settles to a steady state
value $m_{ss}(B)$.
Thus for each value of $B$ one can associate a steady-state value $m_{ss}(B)$ and an initial value $m_{\rm in}=m(B,\gamma=0)$.
At this point we turn to a theoretical analysis whose aim is to understand the steady-state value $m_{ss}(B)$ and to derive an equation for the dependence of $m$ on $\gamma$ for any given value of $\B B$.

\section{The steady state value of the magnetization}
\label{ss}

\begin{figure}
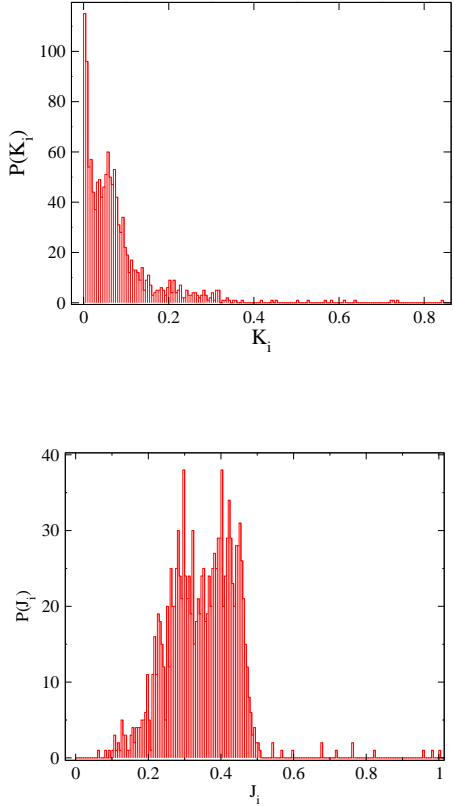

\includegraphics[scale = 0.40]{PlastindmagFig2a.eps}
\vskip 1.2 cm
\includegraphics[scale = 0.40]{PlastindmagFig2b.eps}
\caption{Upper panel: A typical distribution of $K_i$ in our simulations. This data was taken at $\gamma=1.5$ and $B=0.01$ but the distribution is quite insensitive to these parameters, being determined by the atomic randomness. Lower panel:
the distribution of $J_i\equiv J(r_{ij})$ where all the values of $j$ are included for each $i$. The conditions are the same.}
\label{KandJ}
\end{figure}

To develop a theory of the effect displayed in Sect. \ref{effect} we start from the obvious remark that in
an AQS procedure the force (or torque) on any particle is zero before and after every change in strain, and therefore
\begin{eqnarray}
&&-\frac{\partial U}{\partial \phi_i } = \sum_j J(r_{ij}) \sin(\phi_i-\phi_j) +K_i \sin [2(\phi_i-\theta_i)] \nonumber\\&&+B \sin\phi_i=0 \ .
\label{AQS}
\end{eqnarray}
Thus this condition gives rise to $N_A$ coupled nonlinear algebraic equations for the spin coordinates $\phi_i$ given
the instantaneous easy axis directions $\theta_i$. To proceed we will accept Eq. (\ref{ptheta}) and in addition
will make the {\em mean field approximation} that for any finite $\B B$
\begin{eqnarray}
&&\sin(\phi_i-\phi_j) =\sin\phi_i\cos\phi_j - \cos\phi_i\sin\phi_j\nonumber\\
&&= \sin\phi_i\langle \cos\phi_j\rangle  - \cos\phi_i\langle \sin\phi_j\rangle =\sin\phi_i m \ .
\end{eqnarray}
Here we used the fact that $\B B$ is in the $x$ direction such that for any value of $K_i$ we have for the spins $j$ surrounding $i$, $\langle \sin \phi_j\rangle =0$. Next we consider the distributions of $K_i$ and $J(r_{ij})$, Cf. Fig.~\ref{KandJ}.
This distributions are well behaved with a very well defined averages.
Denoting then the number of nearest neighbors by $q$ and the average values of $J( r_{ij})$ and $K_i$ by $\bar J$ and $\bar K$ respectively we find the mean field equations
\begin{equation}
\frac{\bar Jqm+B}{\bar K}\sin\phi_i + \sin [2(\phi_i-\theta_i)] = 0.
\label{achla}
\end{equation}
These equations should be solved together with the supplementary equations (\ref{ptheta}) and (\ref{defm}). The solution
should provide the steady state values of $m_{ss}(B)$ as seen in Fig. \ref{numsol} for given parameters $\bar J$ and $\bar K$ and for any value of $B$.
\subsection{Numerical Solution}
\begin{figure}
\includegraphics[scale = 0.35]{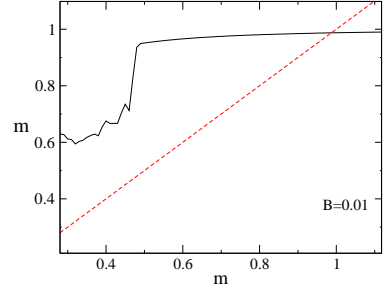}
\caption{An example of the numerical procedure to solve Eq. (\ref{achla}) for $B=0.01$ by
iterations, as shown by the black continuous line. Convergence is obtained when the employed value of $m$ is
returned consistently as explained in the text.}
\label{numsol}
\end{figure}

An easy way to solve Eq. (\ref{achla}) numerically is by graphic methods. Since $\theta_i$ is flatly distributed
in the interval $[0,2\pi]$ we can choose $M$ values of $\{\theta_i=2n\pi/M\}_{n=0}^M$ and then solve for $\phi_i$ using a
Newton-Raphson method starting with $m_1=m_{\rm in}$. The procedure provides values for $\phi_i(m_1)$ from which we compute a new value of $m_2=\langle \cos \phi_i(m_1)\rangle$. Solving again, we get a new set of $\phi_i(m_2)$ and a new value of $m_3$. The procedure is stopped when we get back the same value of $m$, see Fig. \ref{numsol} for an example of this
procedure for $B=0.01$. The converged value of $m$ for this value of $B$ is 0.97, compared to 0.9 from the direct
numerical simulation.
Considering the mean field approximation that is at the basis of Eq. (\ref{achla}) we find this result very satisfactory.

\subsection{Analytic Solution for large $\bar K$ and small $B$}

In this subsection we examine the solution of Eqs. (\ref{achla}) when the effect of random anisotropy is large
(large $\bar K$), the exchange $\bar Jq$ is small and the magnetic field $B$ small. In this case we expect that $\phi_i$ will deviate slightly from $\theta_i$,
$\phi_i=\theta_i+\delta \phi_i$. Linearizing Eq. (\ref{achla}) in $\delta \phi_i$ we find
\begin{equation}
\delta \phi_i = \frac{-(\bar Jqm+B)\sin \theta_i}{(\bar Jqm+B)\cos\theta_i +2\bar K} \ .
\end{equation}
Using this result we can compute directly the magnetization using Eqs. (\ref{defm}) and (\ref{ptheta}),
\begin{equation}
m=\frac{1}{2\pi} \int_{-\pi}^\pi \d\theta \cos[\theta - \frac{-(\bar Jqm+B)\sin \theta}{(\bar Jqm+B)\cos\theta +2\bar K}] \ .
\end{equation}
Finally, since we assumed already that $K$ is large and $B$ is small, we can rewrite this last equation as
\begin{eqnarray}
m&=&\frac{1}{2\pi} \int_{-\pi}^\pi \d\theta \cos[\theta - \frac{-(\bar Jqm+B)\sin \theta}{2\bar K}] \nonumber\\&=& J_1 \left(\frac{-(\bar Jqm+B)}{2\bar K}\right) \,
\label{bessel}
\end{eqnarray}
where $J_1(X)\equiv \frac{1}{2\pi} \int_{-\pi}^\pi \d\theta \cos[\theta - X\sin \theta]$ is the Bessel function of the first kind. As anticipated already from Eq.~(\ref{achla}) the relevant parameter that determines the steady-state value $m_{ss}$ is $X = \frac{(\bar Jqm+B)}{2\bar K}$. For a small value of $X$ we can use the series expansion of the Bessel function
$J_1(X) \approx X/2-X^3/16$ to define and compute the susceptibility $\chi(Jq,K)$,
\begin{equation}
m(\bar Jq,\bar K,B) \approx \chi(\bar Jq,\bar K) B\ , \quad \chi (\bar Jq,\bar K) = \frac{1}{4\bar K-\bar Jq} \ .
\end{equation}

Eq. (\ref{bessel}) can be also solved numerically, and the predicted value of $m$
\begin{figure}
\includegraphics[scale = 0.45]{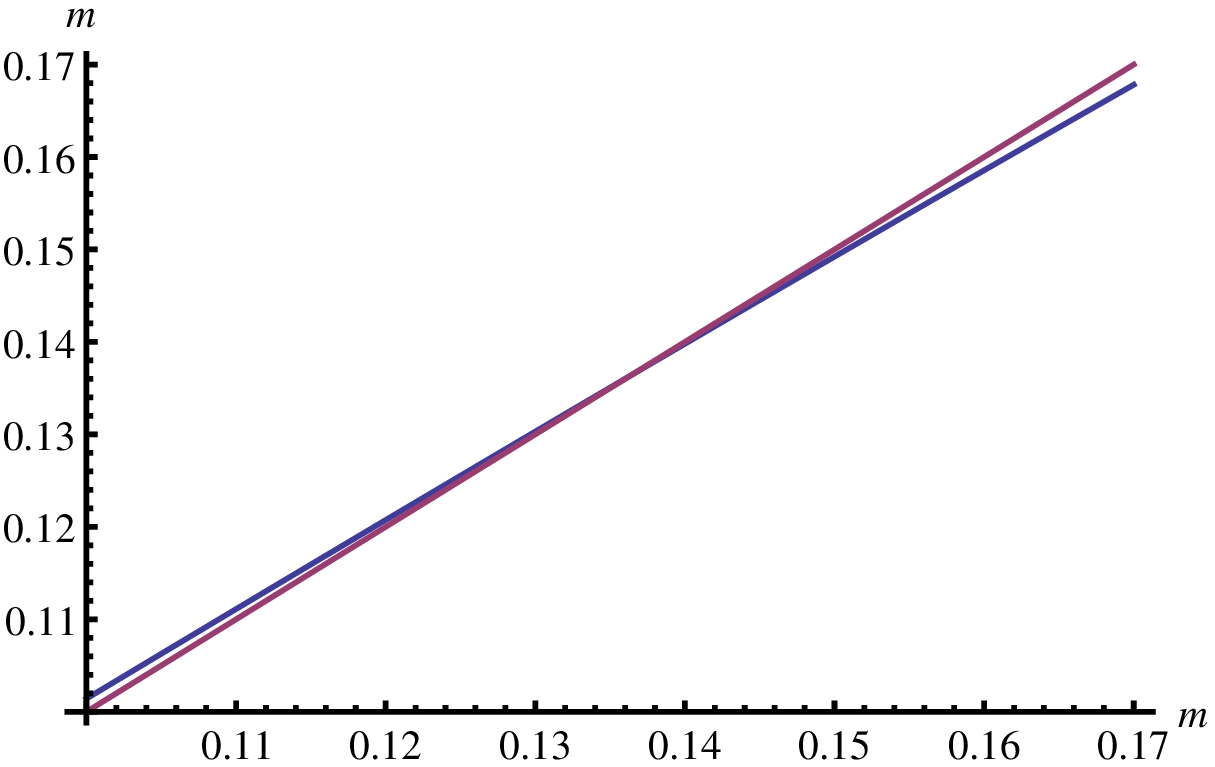}
\includegraphics[scale = 0.45]{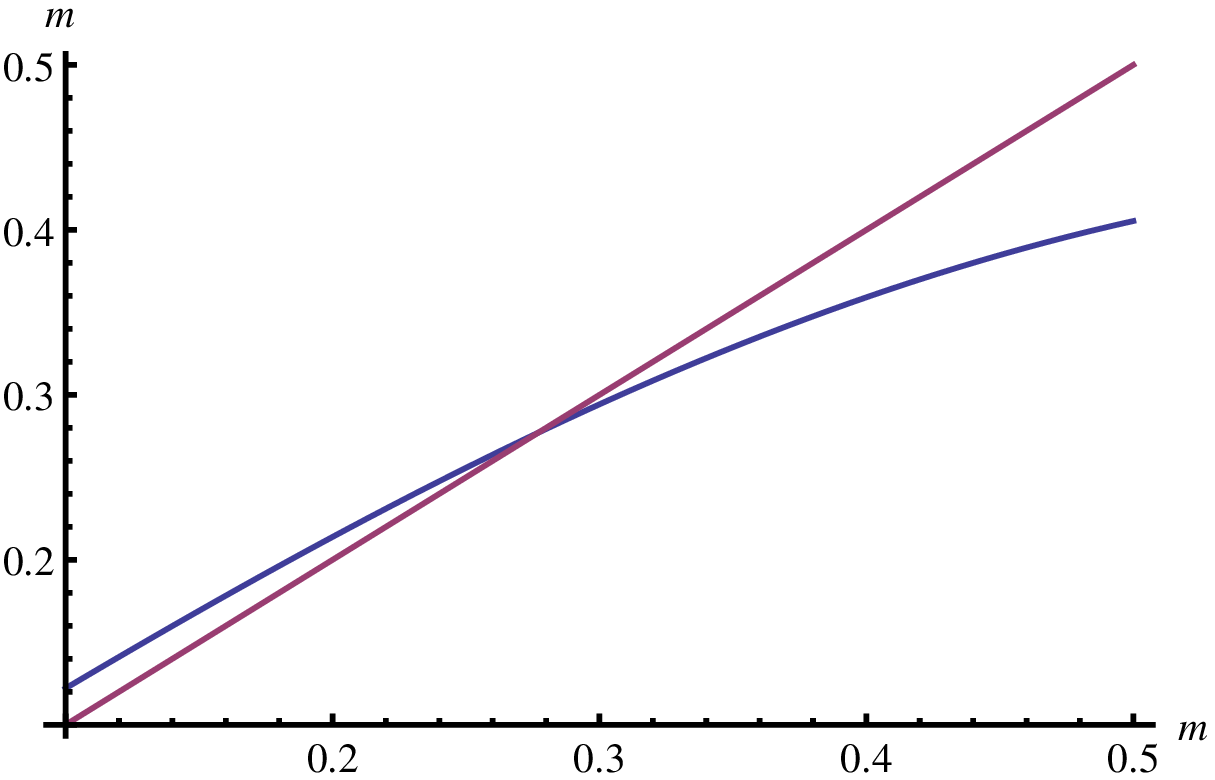}
\caption{Graphic solution of Eq. (\ref{bessel}) for $B=0.001$ and $B=0.01$. The Bessel function intersects with the linear function $m$ at the values $m_{ss}\approx 0.13$ and $m_{ss}\approx 0.28$ respectively.
Note that these results pertain only to $2\bar K\gg \bar Jqm+B$.}
\label{besslsol}
\end{figure}
can be read from the graphic solution as shown in Fig. \ref{besslsol}. For the case $B=0.001$ for which the condition
 $2\bar K\gg \bar Jqm+B$ is valid we get a very good agreement between the value of $m_{ss}$ found here and in the direct
 numerics ($m_{ss}\approx 0.13$ compared to $m_{ss}=0.12$). For the large value of $B=0.01$ the condition $2\bar K\gg \bar Jqm+B$ is not obeyed; the value of the predicted $m_{ss}=0.28$ is not close to the direct numerical value of $m_{ss}=0.9$.
\subsection{General Analytic Solution}

The theory derived above is only valid for  small magnetic fields and weak magnetization. In the next section we derive a nonlinear relaxation equation Eq.~(\ref{3}) for the magnetization as a function of strain. But in order to integrate this equation we will require expressions for the average magnetization for larger magnetic fields and signification magnetization. Further we will see that our expressions for the  magnetization show hysteresis as expected from simulations. We therefore present series solutions in both powers of $X=(\bar Jq m +B)/2\bar K$ and powers of $Y=1/X = 2\bar K/(\bar Jq m +B)$  that converge respectively in the case of large $\bar K$ and small $B$ on the one hand and small $\bar K$ and large $B$ on the other.

\subsubsection{X small}

We now proceed to solve the mean field equation by separation of variables in the form
\begin{equation}
\phi (X, \theta) = \sum_{n=0}^{\infty} X^n a_n(\theta) \ .
\label{achlaa}
\end{equation}
We expect this power series in $X$  to converge for $X<X_c$ (which we estimate below). The coefficients $a_n(\theta)$ are found from the nonlinear equation
\begin{equation}
2 X\sin\phi + \sin [2(\phi-\theta)] = 0.
\label{achlab}
\end{equation}
Equating powers of $X$ we find
\begin{eqnarray}
&&\phi (X, \theta) = \theta -X \sin{\theta} + (X^2/2)\sin{2\theta}\nonumber \\
&&-(X^3/24)(9\sin{\theta}+5\sin{3\theta})+(X^4/2)\sin{2\theta} \label{achlac} \\
&&+ (X^5/640)(-350\sin{\theta}-225\sin{3\theta}+77\sin{5\theta}) + O(X^6)\nonumber
\end{eqnarray}
\begin{figure}
\includegraphics[scale = 0.35]{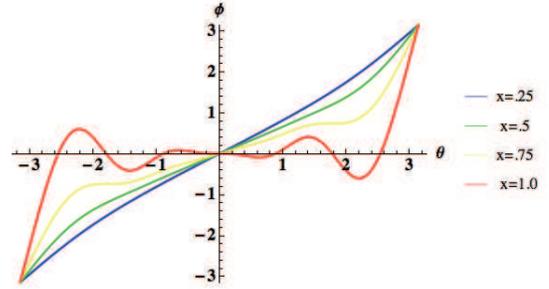}
\caption{Plots of $\phi$ (the equilibrium position of a spin) compared to its local easy axis $\theta$ as $X$ is increased using Eq.~(\ref{achlac}).}
\label{phivst}
\end{figure}
In Fig.~\ref{phivst} we plot $\phi$ the equilibrium position of a spin compared to its local easy axis $\theta$ as $X$ is increased. As can be seen, for $X\ll 1$ the spins are essentially pinned along their easy axes, but as $X$ increases by, for example, increasing the magnetic field $B$, the spins have a tendency to prefer the magnetic axis. This is especially true for easy axes that by chance are already close to the magnetic axis. As $X$ increases further this set increases further, finally only leaving small subsets of spins where the easy axes are oriented close to $\pm \pi$.

Having introduced a series expansion in $X$ we should distinguish between the steady-state value $m_{ss}(X)$ and any intermediate
value of $m$ that is obtained for $\gamma\ll \infty$.
In order to calculate $\langle \cos{\phi}\rangle = m_{ss}(X)$ using Eq.~(\ref{achlac}), we use the series expansion for $\phi (X,\theta )$ with the result
\begin{equation}
\cos{\phi (X, \theta)} = \sum_{n=0}^{\infty} X^n b_n(\theta).
\label{achlad}
\end{equation}
Calculating the first few terms of this expansion we find
\begin{eqnarray}
&&\cos{\phi (X, \theta)} = \cos{\theta } +(X/2) (1 - \cos{2 \theta}) \nonumber \\
&& - (3 X^2/8)(\cos{\theta } - \cos{3\theta }) +(X^3/4)(1-\cos{4\theta} ) \nonumber\\
&&-(5 X^4/128)(10\cos{\theta}-7\cos{3\theta}-3\cos{5\theta})\nonumber \\
&&+ (3X^5/8)(1-\cos{4\theta}) + O(X^6).
\label{achlae}
\end{eqnarray}
From Eq.~(\ref{achlae}) we immediately see that
\begin{equation}
\langle \cos{\phi (X, \theta)} \rangle = \sum_{n=0}^{\infty} X^n \langle b_n(\theta)\rangle = X/2+X^3/4+3 X^5/8 +O(X^7).
\label{achlaf}
\end{equation}
From the inequality $\langle \cos{\phi (X, \theta)} \rangle \le 1$ we see that the expansion in powers only exist for $X<X_{c1}\approx .957$. We therefore now consider solutions in powers of $Y=1/X$.

\subsubsection{Y small}
\noindent The parameter $Y=1/X = 2\bar K/(\bar Jq m +B)$ is small for weak anisotropy and strong magnetic fields. We now proceed to solve the mean field equation for
\begin{equation}
\phi (Y, \theta) = \sum_{n=0}^{\infty} Y^n c_n(\theta)
\label{achlaaa}
\end{equation}
in a power series in $Y$  that will converge for $Y<Y_c$  from the nonlinear equation
\begin{equation}
2 \sin\phi + Y \sin [2(\phi-\theta)] = 0.
\label{achlabb}
\end{equation}
Equating powers of $Y$ we find
\begin{eqnarray}
&&\phi (Y, \theta) =  (Y/2) \sin{2\theta} - (Y^2/4)\sin{4\theta}\nonumber \\
&&-(Y^3/192)(9\sin{2\theta}-35\sin{6\theta})\nonumber\\
&&+(Y^4/32)(2\sin{4\theta} -5\sin{8\theta})\label{achlacc}\\
&&- (Y^5/20480)(50\sin{2\theta}+1575\sin{6\theta}-3003\sin{10\theta})\nonumber
\end{eqnarray}

\begin{figure}
\includegraphics[scale = 0.35]{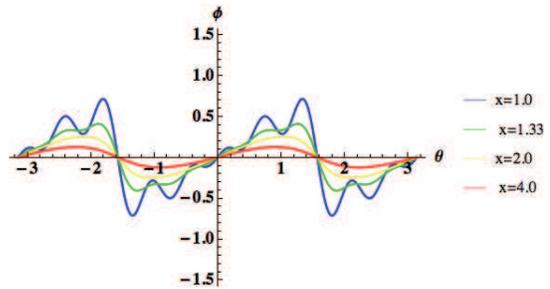}
\caption{Plots of $\phi$ (the equilibrium position of a spin) compared to its local easy axis $\theta$ as $Y$ is increased using Eq.~(\ref{achlacc}).}
\label{phi2}
\end{figure}
In Fig.~\ref{phi2} we plot $\phi$, the equilibrium position of a spin, compared to its local easy axis $\theta$ as $Y$ is increased. As can be seen, for $Y\ll 1$ the spins are essentially pointing along the magnetic axis, but as $Y$ increases by, for example, reducing the magnetic field $B$ the spins have a tendency to prefer the local anistropy easy axis. This is especially true for easy axes that by chance are already close to the magnetic axis. As $Y$ increases further this set increases further, finally only leaving small subsets of spins where the easy axes are oriented close to $\pm \pi$ still preferentially pointing along the magnetic axis.

In order to calculate $\langle \cos{\phi}\rangle = m_{ss}(X)$ using Eq.~(\ref{achlacc}), we use the series expansion for $\phi (Y,\theta )$ with the result
\begin{equation}
\cos{\phi (Y, \theta)} = \sum_{n=0}^{\infty} Y^n d_n(\theta).
\label{achladd}
\end{equation}
Calculating the first few terms of this expansion we find
\begin{eqnarray}
&&\cos{\phi (Y, \theta)} = 1- (Y^2/16) (1 - \cos{4 \theta} \nonumber \\
&& + (Y^3/16)(\cos{2\theta }-\cos{6\theta}) \label{achlaee}\\
&&-(Y^4/3072)(9-100\cos{4\theta}+91\cos{8\theta})\nonumber \\
&&+ (Y^5/64)(2\cos{2\theta}-3\cos{6\theta}+\cos{10\theta}) + O(Y^6).
\nonumber
\end{eqnarray}
From Eq.~(\ref{achlaee}) we immediately see that
\begin{eqnarray}
&&\langle \cos{\phi (Y, \theta)} \rangle = \sum_{n=0}^{\infty} Y^n \langle d_n(\theta)\rangle\nonumber\\ &&= 1- Y^2/16-3Y^4/1024 +O(Y^6).
\label{achlaff}
\end{eqnarray}
From the inequality $\langle \cos{\phi (Y, \theta)} \rangle \ge 0$ we see that the expansion in powers can only exist for $Y<Y_c\approx 3.266$ or $X_{c2}>.306$.

Thus we see from our analysis the magnetization is hysteretic. For $X<X_{c2}\approx .306$ at low magnetic fields only pinned solutions exists. While for $X>X_{c1}\approx .957$ at high magnetic fields only the depinned phase exists. Finally for $X_{c2}<X<X_{c1}$ both phases are possible and the chosen solution will depend on initial conditions.\\

\section{The magnetization $m$ as a function of the strain $\gamma$}
\label{mvsg}

\begin{figure}
\includegraphics[scale = 0.40]{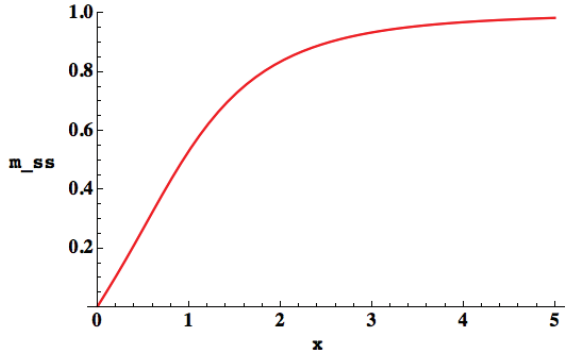}
\caption{Plot of the Pade approximant form for $m_{ss}(X)$ versus $X$ used to integrate Eq.~(\ref{3}). }
\label{pade}
\end{figure}

In this section we derive an approximate differential equation for $m$ as a function of $\gamma$.
To this aim we assume that we know the steady state value of $m$ as a function of the parameter $X = \frac{(\bar Jqm+B)}{2\bar K}$, $m_{ss}(X)$. For a value of $\gamma$ not in the steady state we write
\begin{equation}
m(\gamma+\delta\gamma) = m(\gamma) +\delta m \ .
\label{1}
\end{equation}
To proceed, we realize that the change $\delta m$ occurs only due to plastic events, and these are localized
on a small number of particles $n$, $n\ll N$. We will denote the relative number of particles involved in the
plastic events as $p\equiv n/N$. If $p$ were unity, we would expect that the change $\delta m$ would
be complete, i.e. $\delta m = m_{ss}(X) -m$. Since $p\ll 1$ we estimate
\begin{equation}
\delta m = p [m_{ss}(X)-m] \ .
\end{equation}
Using this estimate in Eq. (\ref{1}) we write
\begin{equation}
\frac{\partial m}{\partial \gamma} \delta \gamma = p [m_{ss}(X)-m] \ .
\label{2}
\end{equation}
Dividing through by $p$ we write
\begin{equation}
\Gamma\frac{\partial m}{\partial \gamma}  = [m_{ss}(X)-m] \ , \quad \Gamma = \delta \gamma/p \ .
\label{3}
\end{equation}
To compare these equations to the simulations we need to show that $\Gamma$ is an intensive parameter, independent of the system size.
This is done in the appendix. In addition,  we employ a Pade' approximant form for $m_{ss}(X)$ that captures both the $X\ll 1$ behavior $m_{ss}(X) \approx X/2 + X^3/4$ and the $X\gg 1$ behavior $m_{ss}(X) \rightarrow 1$. These asymptotic limits are given by the Pade' approximant
\begin{equation}
m_{ss}(X) = \frac{P(X)}{Q(X)} = \frac{(X/2 +X^2/4 + 3 X^3/8)}{(1+ X/2 +X^2/4 + 3 X^3/8)} \ .
\label{3a}
\end{equation}
 This form is plotted in Fig.~\ref{pade}.

\begin{figure}
\includegraphics[scale = 0.40]{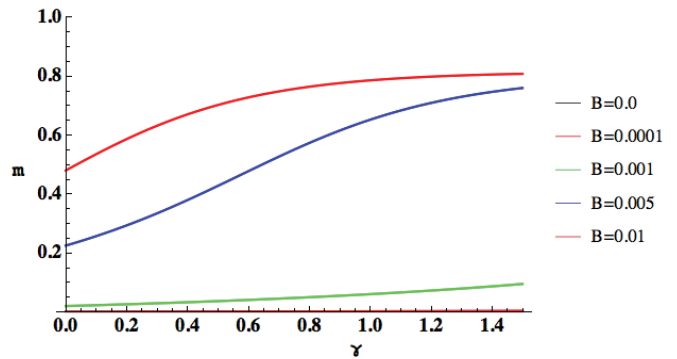}
\caption{Plots of the theoretical values of the magnetization $m$ as a function of the strain $\gamma$ for increasing values of $B$ from $B=0$ to $B=.01$ as shown in the figure. These results are found by integrating Eq.~(\ref{3}) starting from the initial condition  $m(B,\gamma = 0)$.}
\label{plots}
\end{figure}

Eq.~(\ref{3}) is a nonlinear relaxation equation for the magnetization describing its approach to steady state as the material is strained. Using the Pade' approximant form for $m_{ss}(X)$ given by Eq.~\ref{3a} that captures both the small and large $X$ behavior we solved Eq.~(\ref{3}) with the initial condition $m(B,\gamma = 0)$ using the same parameters as in the simulation, namely $Jq\approx .36$, $K\approx .08$. We chose $\Gamma = .15$ to get a qualitative fit with the direct simulations, and solved for various values of $B$ from $B=0$ to $B=.01$. The resultant curves are shown in Fig.~\ref{plots}.
These plots should be compared with the direct numerical simulations in Fig.~\ref{numeffect}. Taking into account the
mean-field approximation that is behind Eq.~(\ref{3}) and the Pade' approximation involved we find the
comparison quite encouraging.

\section{Concluding remarks}
\label{conclude}

In summary, we have presented an interesting effect that is particular to magnetic amorphous solids, showing that
at $T=0$ the plastic events can act as an ``effective noise" that drives the magnetization from initial conditions
to a steady state value. The magnetization is changing in steps that are coincident with the irreversible plastic
events. While the effect itself was discovered numerically, we presented above a mean-field theory that provides
adequate estimates of both the steady-state values of the magnetization and of the trajectories ($m$ vs. $\gamma$)
to get there. The model employed above appears to offer considerable amount of additional interesting physics that calls for
careful study, as will be elaborated in future work.

\appendix
\section{the parameter $\Gamma$}

To show that $\Gamma$ is an intensive parameter we rely heavily on the scaling theory of elasto-plastic steady states that is presented in great details in Ref. \cite{10KLP}. Denoting the steady state mean stress as $\sigma_\infty$, we note that the
value of $\Gamma$ can be obtained by equating the typical increase in elastic energy in the steady state, i.e.
$\Delta U\approx V\sigma_\infty \delta \gamma$ with the typical plastic energy drop $n\epsilon$ where $\epsilon$ is the
plastic energy drop per particle. Writing $n\epsilon= pN\epsilon$ and $V=Nv$ we find $\Gamma\sim \epsilon/(v\sigma_\infty )$.
Here $v$ is the volume per particle. Thus $\Gamma$ is intensive even though neither $\delta \gamma$ nor $p$ are intensive.
In fact, from Ref. \cite{10KLP} we expect both to scale like $N^{-2/3}$. This expectation follows from the scaling
behavior $\Delta U\sim N^\alpha$ and $\delta\gamma\sim N^\beta$ together with the scaling relation $\alpha=1+\beta$ \cite{10KLP}. Finally, since the plastic event is associated with a saddle node bifurcation, we know that
the barrier to instability scales like $\delta \gamma^{3/2}$. In Ref. \cite{10KLP} it is also argued that the barrier
scales like $1/N$ leading finally to $\delta \gamma\sim N^{-2/3}$. Therefore we also find that $\alpha=1/3$ and thus
the participation ratio $p$ also scales like $p\sim N^{-2/3}$. Accordingly $\Gamma$ is intensive. This is crucial for the
comparison of the theory and the simulations as shown above.

\acknowledgments

This work had been supported by an ERC ``ideas" grant STANPAS, the German Israeli Foundation and the
Israel Science Foundation.

\end{document}